\documentclass[aps, twocolumn, a4paper, superscriptaddress]{revtex4}

\usepackage{graphicx}
\usepackage{float}
\usepackage{amsmath, amssymb}
\usepackage{lmodern}
\usepackage[T1]{fontenc}
\usepackage[utf8]{inputenc}
\usepackage[english]{babel}

\begin{document}

\title{Expanding spatial domains and transient scaling regimes in populations with local cyclic competition}

\author{P.P. Avelino}
\affiliation{Instituto de Astrof\'{\i}sica e Ci\^encias do Espa{\c
c}o, Universidade do Porto, CAUP, Rua das Estrelas, PT4150-762 Porto,
Portugal}
\affiliation{Departamento de F\'{\i}sica e Astronomia, Faculdade de
Ci\^encias, Universidade do Porto, Rua do Campo Alegre 687, PT4169-007
Porto, Portugal}
\author{J. Menezes}  
\affiliation{Instituto de Astrof\'{\i}sica e Ci\^encias do Espa{\c
c}o, Universidade do Porto, CAUP, Rua das Estrelas, PT4150-762 Porto,
Portugal}
\affiliation{Escola de Ci\^encias e Tecnologia, Universidade Federal
do Rio Grande do Norte\\
Caixa Postal 1524, 59072-970 Natal, RN, Brazil}
\affiliation{Institute for Biodiversity and Ecosystem Dynamics,
University of Amsterdam, Science Park 904, 1098 XH Amsterdam, The
Netherlands}
\author{B.F. de Oliveira}
\affiliation{Departamento de Física,
Universidade Estadual de Maringá, 
Av. Colombo 5790, 87020-900 Maringá, PR, Brazil}
\author{T. A. Pereira}
\affiliation{Departamento de F\'{\i}sica Te\'orica e Experimental,
Universidade Federal do Rio Grande do Norte, 59078-970 Natal, RN,
Brazil}

\begin{abstract}

We investigate a six-species class of May-Leonard models leading to formation two types of competing spatial domains, each one  inhabited by three-species with their own internal cyclic rock-paper-scissors dynamics. We study the resulting population dynamics using stochastic numerical simulations in two-dimensional space. We find that as three-species domains shrink, there is an increasing probability of extinction of two of the species inhabiting the domain, with the consequent creation of one-species domains. We determine the critical initial radius beyond which these one-species spatial domains are expected to expand. We further show that a transient scaling regime, with a slower average growth rate of the characteristic length scale $L$ of the spatial domains with time $t$, takes place before the transition to a standard $L \propto t^{1/2}$ scaling law, resulting in an extended period of coexistence.
\end{abstract}

\maketitle

\section{Introduction}

There is ample evidence that non-hierarchical interactions between individuals of different species play a crucial role in the development and preservation of biodiversity. Predation, reproduction, and mobility interactions are ubiquitous in nature and constitute a crucial ingredient of most competition models, many of then inspired in the pioneering work by Lotka and Volterra, and May and Leonard \cite{1920PNAS....6..410L,1926Natur.118..558V,May-Leonard}. The so-called rock-paper-scissors model considers three species which cyclically dominate each other  \cite{2002-Kerr-N-418-171,Reichenbach-N-448-1046} (see \cite{2010-Wang-PRE-81-046113,2014-Cianci-PA-410-66,2010-Yang-C-20-2,2017-Park-SR-7-7465,2017-Souza-Filho-PRE-95-062411} for models with additional interactions). Despite its simplicity, it has been successful in reproducing crucial dynamical features of some biological systems composed of three species with cyclic selection interactions \cite{lizards,2002-Kerr-N-418-171,bacteria} (see \cite{2014-Szolnoki-JRSI-11-0735,2018JPhA...51f3001D} for recent reviews).

In \cite{Avelino-PRE-86-031119, Avelino-PRE-86-036112}, a broad family of spatial stochastic May-Leonard models with an arbitrary number of species has been introduced, thus  generalizing the standard rock-paper-scissors model (see also \cite{PhysRevE.76.051921,Peltomaki2008,Szabo2008,Hawick2011,Hawick_2011}). Some of these models were shown to give rise to complex spatial structures, which may include spirals with an arbitrary number of arms \cite{Avelino-PRE-86-036112}, interfaces (which may themselves develop a non-trivial internal dynamics \cite{Avelino-PRE-89-042710,PhysRevE.96.012147,z2}) and strings with or without junctions \cite{Avelino-PLA-378-393,2017-Avelino-PLA-381-1014}. On the other hand, the scaling laws governing the dynamics of such systems may also be very diverse \cite{Avelino-PRE-86-036112,PhysRevE.96.012147} (see also \cite{2017-Maynard-NEE-6-01256,2017-Pascual-Garcia-NC-8-14326,doi:10.1080/17429145.2011.556262} for a discussion of the role of partnerships in the coexistence of biological systems).

In this paper we consider the population dynamics, in two spatial dimensions, for a particular $6$-species sub-class of the more general family of May-Leonard models with an arbitrary number of species introduced in \cite{Avelino-PRE-86-031119, Avelino-PRE-86-036112}. We shall investigate in detail a particular property of this sub-class of models associated with the formation of single-species spatial domains during the final stages of the collapse of three-species spatial domains, and their subsequent growth if their initial size is above a certain critical radius. The potential impact on the scaling law describing the time evolution of the characteristic length scale of the spatial domains will also be investigated.

The outline of this paper is as follows. In Sec. \ref{sec2} we introduce the class of models investigated in the present paper. The results of a spatial stochastic numerical simulation of the corresponding population dynamics are presented and analysed in Sec. \ref{sec3}. In Sec. \ref{sec4} we investigate, both analytically and numerically, a novel feature present in these simulations: the expansion spatial domains occupied by a single species. In particular, a critical initial radius, beyond which spherically symmetric spatial domains are expected to expand, is determined using spatial stochastic numerical simulations, and the result is compared with the analytical expectations. The impact of the expansion of single-species spatial domains on the evolution of the population is studied in Sec. \ref{sec5}, with a particular emphasis on the scaling of the characteristic length scale with time. Finally we conclude in Sec. \ref{sec6}.

\begin{figure}
	\centering
		\includegraphics{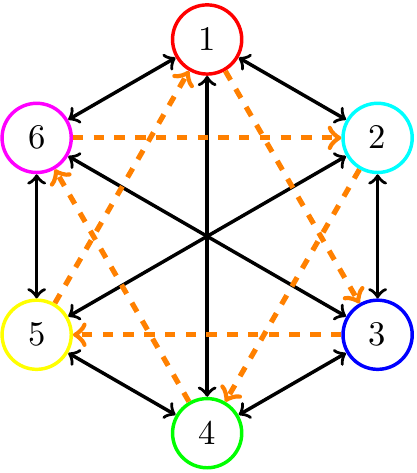}
	\caption{Predation rules in our model: bidirectional predation (solid arrows) and cyclic predation (dashed arrows). The predation  interaction probability of species $i$ with species $i \pm 1$, and $i \pm 3$ is equal to $p$ (solid arrows). The predation interaction probability of species $i$ with species $i+2$ is equal to $p_{PRS}$ (dashed arrows). Except for the labelling of the different species, this figure is invariant under rotation by an angle of $2 \, k\,\pi/6$, where $k$ is an integer, thus leading to a $Z_6$ symmetry.}
	\label{fig1}
\end{figure}
\begin{figure*}
	\centering
		\includegraphics{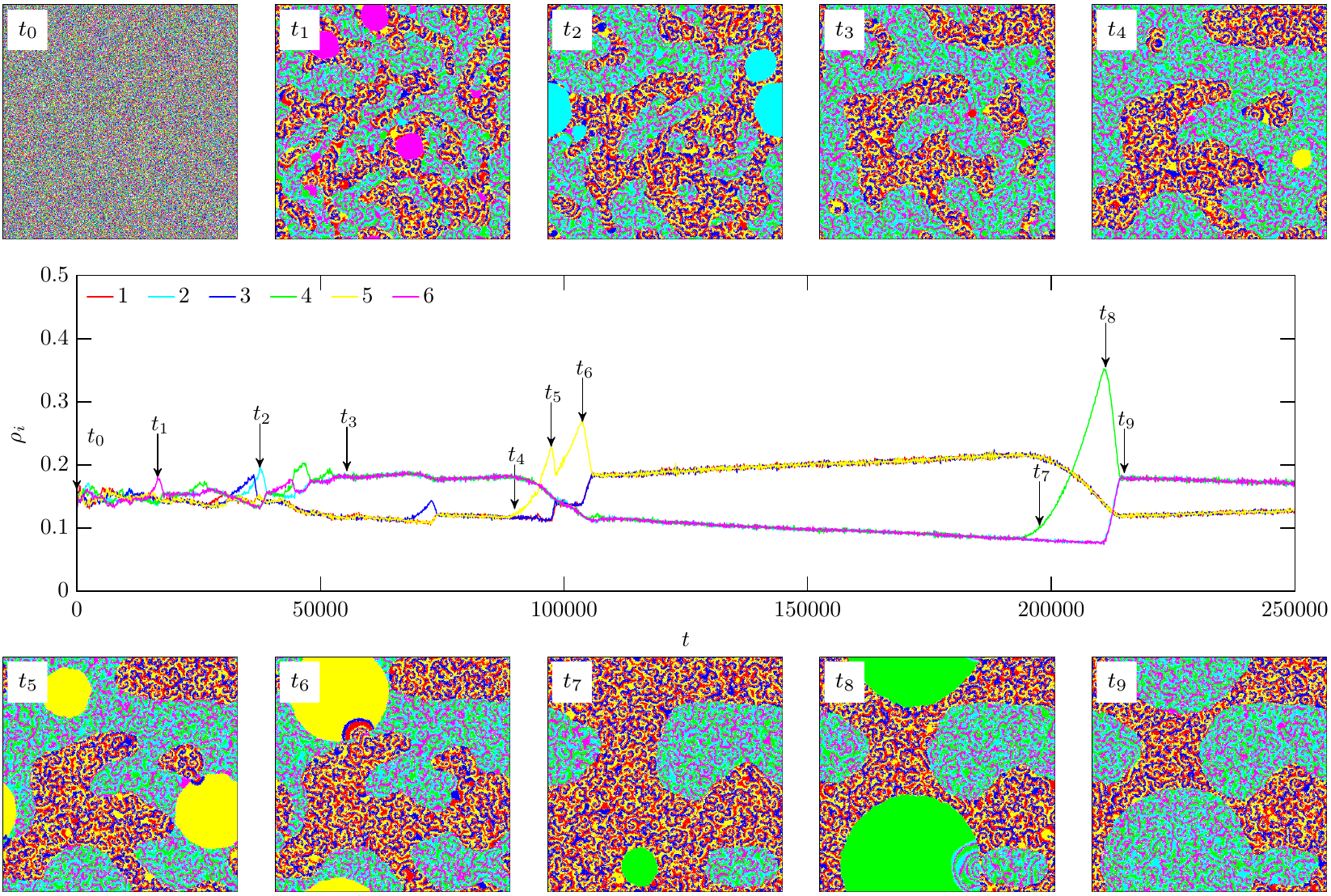}
	\caption{The upper and lower panels show snapshots of the spatial distribution of the different species on the $2000^2$ lattice at various moments for one single realization of our model (with $p=0.25$, $p_{PRS}=p$, $m=0.5$, $r=0.25$), whereas the central panel shows the fractional density of the different species $\rho_i$ for the entire timespan of the simulation. The arrows highlight the instants of time corresponding to the snapshots indicated in the lower and upper panels.}
	\label{fig2}
\end{figure*}

\section{Models \label{sec2}}

In this paper we shall consider a particular sub-class of the more general family of May-Leonard models with an arbitrary number of species ($N$) introduced in \cite{Avelino-PRE-86-031119, Avelino-PRE-86-036112}. In these models individuals of various species are distributed on a square lattice with ${\mathcal N}$ sites and periodic boundary conditions. Each site may be either empty or occupied by a single individual. The different species are labelled by the number $i$ (or $j$), with $i,j=1,...,N$, and empty sites shall be denoted by $\otimes$. The number of individuals of the species $i$ will be denoted by $I_i$ and the number of empty sites by $I_\otimes$. The possible interactions are: predation 
\begin{equation}
i\ j \to i\ \otimes\,, \nonumber
\end{equation}
mobility 
\begin{equation}
 i\ \odot \to \odot\ i\,, \nonumber
\end{equation}
and reproduction 
\begin{equation}
 i\ \otimes \to ii\,, \nonumber
\end{equation}
where $\odot$ represents either an individual of any species or an empty space. 

Here, we shall consider models with $6$ species ($N=6$). The mobility and reproduction interactions occur with probabilities $m$, and $r$, respectively (the same for all species), and the predation probability may be either $p$ or $p_{RPS}$ according to the scheme presented in  Fig.~\ref{fig1} (solid black and dashed orange arrows represent the predation interaction probabilities $p$ and $p_{RPS}$, respectively). Note that the mobility, reproduction, and predation probabilities are assumed to be independent of the position of the individuals in the simulation box. When labelling the species we use modular arithmetic, where numbers wrap around upon reaching $1$ or $N$ (the numbers $i$ and $j$ represent the same species whenever $i=j \, {\rm mod} \, N$, where $\rm mod$ denotes the modulo operation). Except for the labelling of the different species, Fig.~\ref{fig1} is invariant under rotation by an angle of $2 \, k\,\pi/6$, where $k$ is an integer, thus leading to a $Z_6$ symmetry. 

In our model, at each simulation step, the algorithm randomly selects an occupied site to be the active one, randomly chooses one of its four neighbour sites to be the passive one, and randomly picks an interaction to be executed by the individual at the active position ---  in this paper we use the von Neumann neighbourhood (or 4-neighbourhood) composed of a central cell (the active one) and its four non-diagonal adjacent cells. If the interaction cannot be performed (for example, if the passive is an empty site and a predation interaction is picked), the three steps are repeated until a possible interaction is selected. $\mathcal{N}$ successive interactions are completed in one generation time (our time unit).

\section{Population dynamics \label{sec3}}

In this section we shall consider the results of a spatial stochastic numerical simulation with random initial conditions, where at each site an individual of any of the $6$ species or an empty site was selected with a uniform discrete probability of $1/7$ at the beginning of the simulation. Figure ~\ref{fig2} presents results obtained from a single realization of a $2000^2$ lattice numerical simulation of our model assuming $p=0.25$, $p_{PRS}=p$, $m=0.5$, $r=0.25$. The upper and lower panels show snapshots of the spatial patterns at various instants of time:  $t_0=0$, $t_1=16600$, $t_2=37555$, $t_3=55420$, $t_4=89870$, $t_5=97360$, $t_6=103784$, $t_7=197600$, $t_8=211161$, and $t_9=215000$. The central panel shows the  density of the different species 
\begin{equation}
\rho_i=I_i/\mathcal{N}\,,
\end{equation}
for the entire timespan of the simulation: $\Delta t=250000$. The arrows highlight the instants of time corresponding to the snapshots shown in the lower and upper panels, and the colors follow the scheme depicted in Fig. ~\ref{fig1}. The video in \cite{video1} shows the evolution of the spatial patterns for the entire timespan of the simulation.

The snapshot at $t=t_0=0$ depicts the random initial conditions. After an initial stage, essentially two types of spatial domains appear --- a spatial domain being defined as a connected spatial patch dominated by individuals belonging to a single partnership (either $\{1,3,5\}$ or $\{2,4,6\}$). This occurs because mutual predation  takes place between the groups of species 
\begin{equation}
\{1, 3, 5\}, \qquad \{2, 4, 6\}\,. \nonumber
\end{equation}
As a result, two enemy partnerships are formed, as shown in the snapshot taken at $t_1$. Two different species are said to be enemies  if they have bidirectional predation interactions between them --- the species connected by double-headed arrows in Fig. ~\ref{fig1}. The two groups of species $\{1,3,5\}$ and $\{2,4,6\}$ shall be referred to as enemy partnerships in the remainder of the paper.

However, partners do not live peacefully within the spatial domains. On the contrary, they interact with a cyclic predation rule (dashed lines of Fig. ~\ref{fig1}), creating a local rock-paper-scissors dynamics. Although spiral waves travel across the spatial domains, they do not cross the boundaries due to the mutual predation between the members of distinct partnerships. The dynamics of the interfaces between three-species spatial domains is curvature dominated, as described in detail in \cite{Avelino-PRE-86-031119}, with their velocity being roughly proportional to their curvature.

However, Fig. \ref{fig2} shows that the extinction of one of the species in the final stages of the collapse of three-species spatial domains can lead to the formation of spatial domains occupied by a single species, which may then be able to expand. For example, the snapshots taken at $t=t_1$ and $t=t_2$ depict small single-species spatial domains of individuals of species $6$ and $2$, respectively (note that no single-species spatial domain is present in the snapshot taken at $t=t_3$). 

\begin{figure*}
	\centering
		\includegraphics{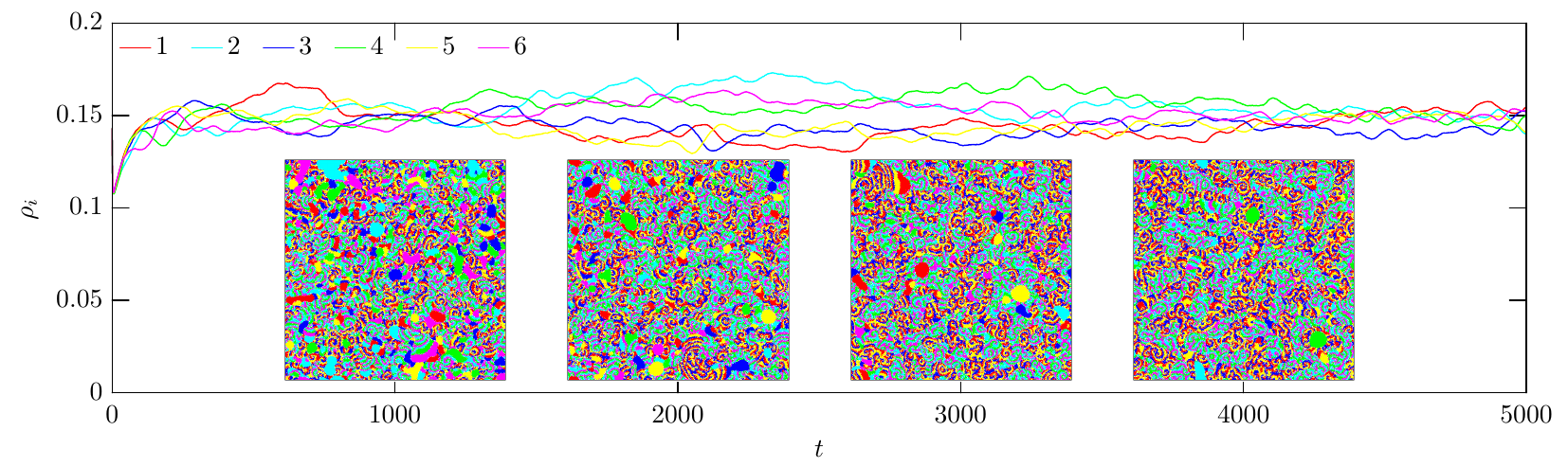}
	\caption{The initial stage of the single simulation presented in Fig. \ref{fig2}. The four snapshots show the spatial distribution of the individuals at the instants of time indicated in the figure. As in Fig. \ref{fig2}, the solid lines represent the evolution of the density of individuals of the different species as a function of time.}
	\label{fig3}
\end{figure*}

In general, outside a single-species spatial domain, individuals have to deal with a cyclic predation among partners besides competing with the enemy partnership. As a consequence, individuals from the single-species spatial domain may have the chance of invading the enemy partnership. This is responsible for the expansion of the single-species spatial domains, as shown in the snapshots taken at $t=t_4$ and $t=t_5$ (we shall quantify, in the following section, the condition for a spherically symmetric single-species spatial domain to be able to expand). As a single-species spatial domain grows larger, it may reach regions occupied by individuals of their original partnership. When this happens it is immediately invaded by a spiral wave front, as depicted in the snapshot taken at $t=t_6$. The encroaching then causes the disappearance of the single-species spatial domain and, consequently, the decrease of the density of the corresponding species.

In summary, the three main factors affecting the dynamics are: I. the curvature dominated dynamics of interfaces separating spatial domains with three-species enemy partnerships; II. the spiral wave fronts inside the spatial domains and the interference between them, which play a crucial role in the creation of small spatial domains occupied by a single species; III. the growth of one-species spatial domains, if their initial size is large enough.

For large $t$, the characteristic size $L$ of the spatial domains increases and, as a consequence, the rate of formation of single-species spatial domains decreases (a rigorous definition of $L$ will be given in Sec. \ref{sec5}). However, once they emerge, they have room to expand further on the grid. As shown in the snapshots taken at $t=t_5$, $t=t_6$ and $t=t_8$, these single-species spatial domains can then grow to become with a characteristic size $L$ comparable to the one of the three-species spatial domains. The vast territorial invasion of single-species areas for large $t$ leads to an increasingly abrupt variation of the densities shown in the central panel of Fig. \ref{fig2}. 

On the other hand, for small $t$ the average size of the spatial domains is tiny and single-species spatial domains are present throughout the whole lattice. Figure 3 depicts the initial stage of the single simulation shown in Fig. 2. As in Fig. \ref{fig2}, the solid lines represent the evolution of the density of individuals of the different species as a function of time. The snapshots, taken after $1000$, $2000$, $3000$, and $4000$ generations, show a fast decrease of the number of single-species spatial domains.

The larger three-species spatial domains are, the longer they take to collapse (the collapse time $t_c$ being roughly proportional to their initial area $t_c \propto L^2$ for a curvature dominated evolution \cite{Avelino-PRE-86-031119}). Hence, one expects the number of collapses per unit area per unit time to scale with $1/(L^2 t_c) \propto L^{-4}$, which is roughly consistent with our numerical results. Since, one-species spatial domains are formed at the end stages of collapse of three-species spatial domains, the number of one-species domains formed per unit time per unit volume is also roughly proportional to $L^{-5}$. This rough estimate implies that the formation of single-species domains was much more frequent at early than at late times, as the results of the simulations confirm.

\section{Expansion of single-species spatial domains \label{sec4}}

If $p_{PRS}=0$ then the local cyclic competition is not present and, therefore, all the empty sites are at the spatial domains' borders (the average density of individuals being the same inside all the spatial domains). In this case, the associated population dynamics has been shown to be curvature driven, analogously to that of a wide variety of material systems, including foam coarsening and grain growth, with the characteristic size $L$ of the spatial domains growing proportionally to $t^{1/2}$  \cite{Avelino-PRE-86-031119}. If $p_{PRS}=0$ a circular interface of thickness
\begin{equation} 
\delta=R_{out}-R_{in}\,,
\end{equation}
has always a tendency to collapse because the average number of predation interactions with the enemy partnership performed, per unit of time, by individuals of the outer spatial domain --- proportional to the external radius $R_{out}$ --- is larger than those performed by individuals of the inner one --- proportional to the internal radius $R_{in}$ (the interface thickness $\delta$ is a function of the parameters $p$, $m$ and $r$).

\begin{figure}
	\centering
\includegraphics{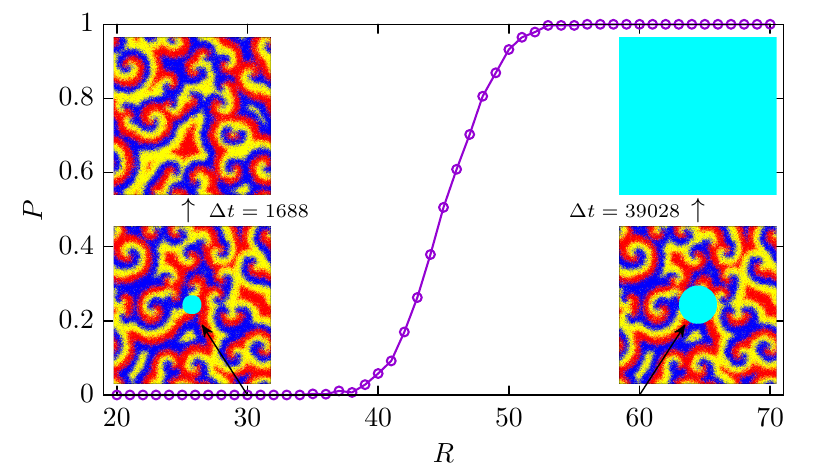}	
	\caption{The probability $P$ that the whole lattice becomes dominated by the species $2$ (initially confined to a circular spatial domain of radius $R$) as a function of $R$, assuming that $p=0.25$, $p_{PRS}=0.25 \, p$, $m=0.5$, $r=0.25$. The results for each $R$ are taken from an average over $1000$ simulations, considering different initial conditions for the outer spatial domain containing species ${1,3,5}$. The critical radius, defined by $P(R_c)=1/2$ is approximately equal to $R_c = 45$ (grid points). The left and right insert panels show two snapshots of runs with initial radius $R = 30$ and $R = 60$, respectively. The times required for the circle to collapse (left inset panels) or to invade all the territory (right inset panels) are displayed between snapshots  (the lower and upper inset panels represent the initial and final configurations).}
	\label{fig4}
\end{figure}

However, if $p_{PRS} > 0$ this is not necessarily the case. One-species spatial domains have a larger density of individuals than three-species spatial domains because the density of individuals inside three-species spatial domains is reduced due to local cyclic competition. Here, we shall demonstrate that this effect may more than compensate the impact of the spatial domain curvature, provided that certain conditions are satisfied.

Consider a circular one-species spatial domain surrounded by a three-species one. Let us denote the density of empty sites away from the borders in three-species spatial domains by 
\begin{equation} 
\rho^{\star}_{\otimes}=I^{\star}_{\otimes}/\mathcal{N}\,.
\end{equation}  
Note that in one-species spatial domains the corresponding density of empty sites is equal to zero. If $p_{PRS} > 0$ then $\rho^{\star}_{\otimes} > 0$ (the in-team predation in the outer three-species spatial domain reduces the number of individuals available to compete with individuals of the enemy partnership). In this case, the average number of predation interactions with the enemy partnership performed, per unit of time, by individuals of the inner one-species spatial domain is still proportional to $R_{in}$, but those performed by individuals of the outer one-species spatial domains becomes proportional to $R_{out} (1-\rho^{\star}_{\otimes})$. One may then define the critical radius as the value of 
\begin{equation} 
R=(R_{out}+R_{in})/2\,,
\end{equation}
for which the average rate of predation interactions with the enemy partnership performed by individuals of the inner one-species and the outer three-species spatial domains are equal, that is \begin{equation} 
R_{in}=R_{out} (1-\rho^{\star}_{\otimes})\,.
\end{equation} 
The critical radius is then equal to
\begin{equation}
R_c = \frac{\delta}{2} \left(\frac{2}{\rho^{\star}_{\otimes}}-1\right)\,. \label{rc}
\end{equation}
If $\rho^{\star}_{\otimes} \ll 1$ then $R_c \propto \delta (\rho^{\star}_{\otimes})^{-1}$. If $R>R_c$ [$R_{in}>R_{out} (1-\rho^{\star}_{\otimes})$] the effect of the larger density of individuals in the inner one-species spatial domain is (on average) the dominant dynamical effect and the circular spatial domain is expected to expand while if $R < R_c$ [$R_{in}<R_{out} (1-\rho^{\star}_{\otimes})$] the dynamics is (on average) curvature dominated and the spatial domain is expected to collapse.

\begin{figure}
	\centering
		\includegraphics{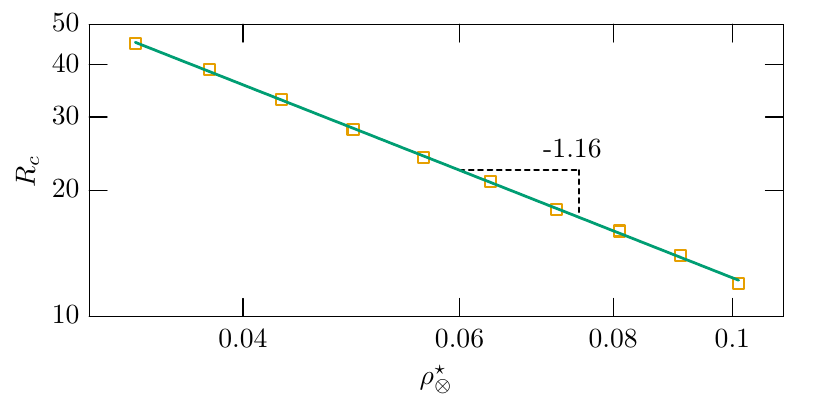}
		\includegraphics{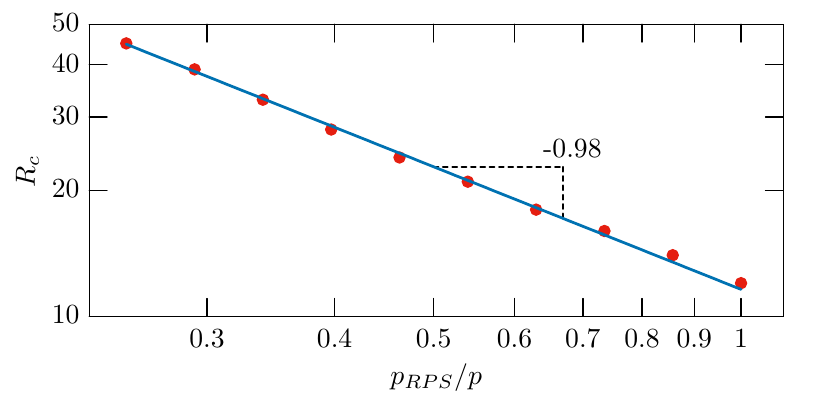}
	\caption{The critical radius $R_c$ as a function of the density $\rho^{\star}_{\otimes}$ of empty sites in the outer spatial domain (top panel) and of $p_{RPS}$ (bottom panel), assuming that $p=0.25$, $m=0.5$, and $r=0.25$. The best fits represent the power laws $R_c \propto (\rho^{\star}_{\otimes})^{\beta_2}$ and $R_c \propto p_{PRS}^{\beta_1}$, with exponents $\beta_1=-1.16$ and $\beta_2=-0.98$, respectively.}
	\label{fig5}
\end{figure}

Figure ~\ref{fig4} considers the collapse (left inset panels) or expansion (right inset panels) of a single-species spatial domain. It displays the probability $P$ that the whole lattice becomes dominated by the species $2$ (initially confined to a circular spatial domain of radius $R$) as a function of $R$, assuming that $p=0.25$, $p_{PRS}=0.25 \, p$, $m=0.5$, $r=0.25$. The results for each $R$ are taken from an average over $1000$ simulations, considering different initial conditions for the outer spatial domain containing species $\{1,3,5\}$. The one-sigma uncertainty in the value of $P$, at each point, may be estimated as $(P(1-P)/1000)^{1/2}$ (with a maximum of $0.016$ for $P=0.5$). The critical radius, defined by $P(R_c)=1/2$ is approximately equal to $R_c = 45$ (grid points). The left and right inset panels show two snapshots of runs with initial radius $R = 30$ and $R = 60$, respectively. The times required for the circle to collapse (left inset panels) or to invade all the territory (right inset panels) are displayed between snapshots (the lower and upper inset panels represent the initial and final configurations).

Figure ~\ref{fig5} shows the value of the critical radius as a function of the density of empty sites $\rho^{\star}_{\otimes}$ in the outer spatial domain (top panel) and of $p_{RPS}/p$ (bottom panel), assuming that $p=0.25$, $m=0.5$, and $r=0.25$. The best fits represent the power laws $R_c \propto (\rho^{\star}_{\otimes})^{\beta_1}$ (top panel) and $R_c \propto p_{PRS}^{\beta_2}$ (bottom panel), with exponents $\beta_1=-1.16$ and $\beta_2=-0.98$, respectively. The result obtained for $\beta_1$ is in reasonable agreement with the analytical expression in Eq. (\ref{rc}) which gives $R_c \propto (\rho^{\star}_{\otimes})^{-1}$ for $\rho^{\star}_{\otimes} \ll 1$. Note that since relative importance of the constant term in Eq. (\ref{rc}) is smaller than $5\%$ for $\rho^{\star}_{\otimes} < 0.1$, it does not have a strong effect on the scaling exponent given in Fig. ~\ref{fig5} (top panel). On the other hand, Eq. (\ref{rc}) with $\delta$ independent of $p_{PRS}$ should be taken as as a rough approximation valid for $p_{PRS} \ll 1$.

Given that the average density of empty sites in the outer spatial domain is expected to be roughly proportional to $p_{PRS}$, $\beta_2$ is also expected to be close to $-1$, which is in agreement with our numerical results. Note that the empty sites in the outer spatial domain are created due to the cyclic competition between the three species which populate that domain. An empty space is created whenever a predation interaction occurs in the outer spatial domain. Taking into account that predation is selected with probability $p_{PRS}$, one expects the average density of empty sites in the outer domain to be roughly proportional to $p_{PRS}$ (an expectation that has also been confirmed numerically).

\begin{figure}
	\centering
		\includegraphics{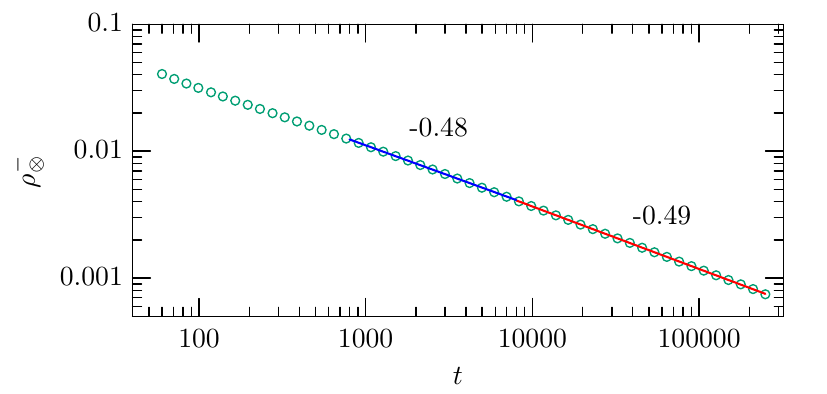}
		\includegraphics{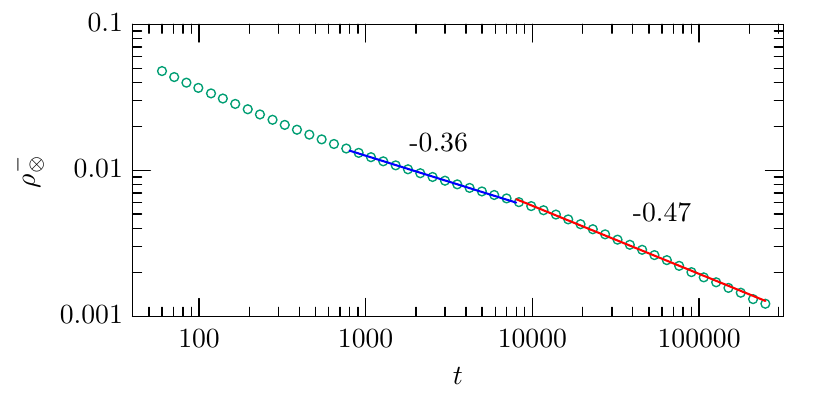}
		\includegraphics{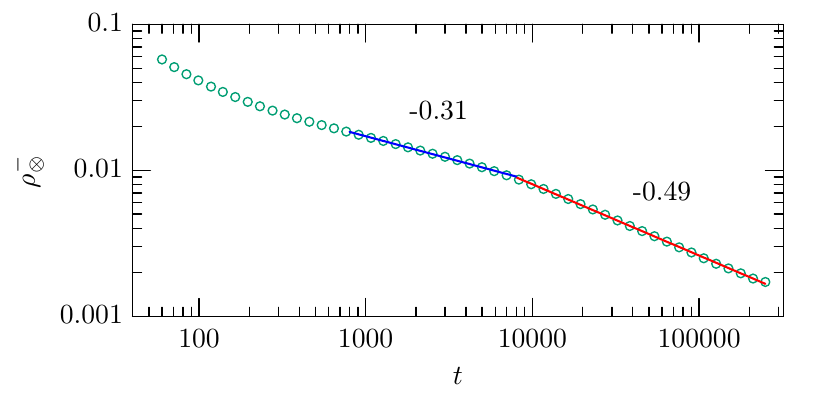}
	\caption{Evolution of the density $\rho^{-}_{\otimes}$ of empty sites between the enemy partnership spatial domains as a function of the simulation time (or, equivalently, the number of generations) assuming that $p=0.25$, $m=0.5$, $r=0.25$, and either $p_{PRS}=0$ (top panel), $p_{PRS}=0.25 p$ (middle panel) or $p_{PRS}=p$ (bottom panel). Both the points and the scaling exponents were obtained from an average over a set of $100$ realizations with different initial conditions. The empty sites between spiral arms inside the spatial domains have not been considered in this analysis.}
	\label{fig6}
\end{figure}

\section{Scaling laws \label{sec5}}

Empty sites appear in the simulations both due to the mutual predation interactions between the enemy partnerships $\{1, 3, 5\}$ and  $\{2, 4, 6\}$, and as a consequence of local cyclic predation interactions taking place mainly between spiral arms. The density of empty sites associated to the interaction between enemy partnerships is given by 
\begin{equation}
\rho^{-}_{\otimes} = I^{-}_{\otimes}/\mathcal{N}\,, 
\end{equation}
where $I^{-}_{\otimes}$ is the total number of empty sites generated at the borders of the spatial domains. The characteristic length $L$ of the spatial domains may be defined as the ratio between area of the square lattice box and the total interface length $\ell$ ($L=\ell^{-1}$ if both $L$ and $\ell$ are expressed in units of the length of the square box). Given that the interface thickness is essentially fixed in the entire grid, the total interface length $\ell$ is proportional to the total number of empty sites at the borders of the spatial domains $I^{-}_{\otimes}$. Hence $L \propto (I^{-}_{\otimes})^{-1} \propto (\rho^{-}_{\otimes})^{-1}$. To distinguish the empty sites associated to predation interactions between the enemy partnerships $\{1, 3, 5\}$ and  $\{2, 4, 6\}$ from the empty sites associated to local  rock-paper-scissors cyclic predation interactions, the four grid sites surrounding each empty site are checked:  if individuals of different partnerships are observed, the empty space is assumed to be associated to the corresponding interface separating enemy partnerships. Otherwise, the empty site is assumed to be due to in-team predation interactions.

Figure ~\ref{fig6} shows the evolution of the density $\rho^{-}_{\otimes}$ of empty sites between the enemy partnership spatial domains with the simulation time (or, equivalently, the number of generations) for a model with $p=0.25$, $m=0.5$, $r=0.25$, and either $p_{PRS}=0$ (top panel), $p_{PRS}=0.25 p$ (middle panel) or $p_{PRS}=p$ (bottom panel). Figure ~\ref{fig6} also displays the scaling exponents $\alpha$, defined by $\rho^{-}_{\otimes} \propto t^\alpha$, associated to two distinct dynamical stages. Both the points and the scaling exponents were obtained from an average over a set of $100$ simulations with different initial conditions. The time intervals $[t_1,\,t_2]=[800,8000]$ and $[t_2,\,t_3]=[8000,\,250000]$ were used to compute the two exponents (the initial stages of the simulation were discarded in this computation). Two different straight lines of the form $\ln \rho^{-}_{\otimes} = \alpha \ln t + C$, where $\alpha$ and $C$ are real parameters, were fit to the data in the time intervals $[t_1,\,t_2]$ and $[t_2,\,t_3]$, respectively. The corresponding scaling exponents $\alpha$ were determined with a $\chi^2$ minimization in $(\ln t, \ln \rho^{-}_{\otimes})$ space. For $p_{PRS}=0$ both exponents, computed using these time intervals, are within $5\%$ of the analytical expectation ($\alpha=-1/2$) for a curvature dominated population dynamics. In order to test the robustness of our results with respect to such splitting, we have also computed the exponents considering time intervals in which $t_1$ and $t_3$ were held fixed and $t_2$ was taken as a free parameter. The parameter $t_2$ has been computed, in each case, with the same  $\chi^2$ minimization described before --- in this case considering the time interval  $[t_1,\,t_3]$ and including $t_2$ as an extra parameter (the corresponding time intervals being $[t_1,\,t_2]=[800,\,4980]$ and $[t_2,\,t_3]=[4980,\,250000]$ for $p_{PRS}/p=1$, and $[t_1,\,t_2]=[800,\,19443]$ and $[t_2,\,t_3]=[19443,\,250000]$ for $p_{PRS}/p=0.25$). The change in the value of the exponents was found to be less than $5\%$. 

As expected, if $p_{PRS}=0$ both scaling exponents are close to $-0.5$  which is the result expected if the dynamics is curvature dominated. This is indeed the case since for $p_{PRS}=0$ the model is equivalent to a two species May-Leonard model having mutual predation with probability $p$ (see, for example, \cite{z2}). However, for $p_{PRS}=0.25 \, p$ (middle panel) or $p_{PRS}=p$ (bottom panel) the two scaling exponents are quite different, the later one being again close to the $\alpha=-0.5$ regime usually associated to a curvature dominated dynamics. In this case, however, due to the emergence of single-species spatial domains the dynamics is never fully dominated by curvature even in the $\alpha \sim -0.5$ regime. What the scaling results show is that, in this regime, the average impact of the single-species spatial domains on the evolution of the characteristic length scale $L$ is small. 

\begin{figure}
	\centering
		\includegraphics{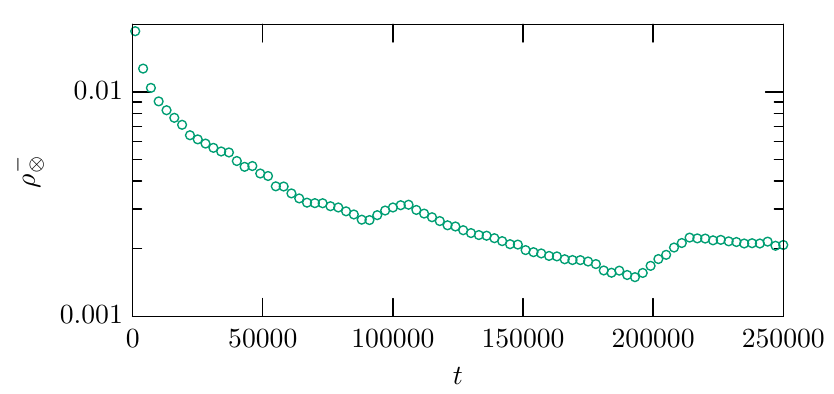}
	\caption{The density $\rho^{-}_{\otimes}$ of empty sites between enemy partnership spatial domains as a function of the simulation time (or, equivalently, the number of generations) for a single realization, assuming that $p=0.25$, $m=0.5$, $r=0.25$, and $p_{PRS}=p$.}
	\label{fig7}
\end{figure}

Figure ~\ref{fig6} shows that the $L \propto t^{1/2}$ regime is preceded by a slower evolution stage where the scaling exponent departs significantly from $-0.5$, ($\alpha=-0.36$ and $\alpha=-0.31$, for $p_{prs}=0.25 \, p$ and $p_{prs}=p$, respectively). In this phase, the higher $p_{RPS}$ is the more the scaling exponent deviates from $-0.5$, which results in an extended period of coexistence. Also note that sharp variations on the evolution $\rho^{-}_{\otimes}$ associated with the expansion of single-species spatial domains are averaged out in Fig. 6 but are expected to be present in any single realization of our model.

Figure ~\ref{fig7} depicts the density of empty sites between enemy partnership spatial domains for the single realization shown in Fig. 2. The positive variations of $\rho^{-}_{\otimes}$ in Fig. 7 are associated to the expansion of single-species spatial domains occurring between the time intervals $[t_4,t_6]$ and $[t_7,t_8]$, depicted in the corresponding snapshots shown in Fig. 2.

\section{Conclusions \label{sec6}}

In this paper we investigated the formation and subsequent growth of single-species spatial domains in six-species class of May-Leonard models. These models naturally lead to the formation of spatial patterns with two types of spatial domains containing individuals from two different three-species partnerships. On the other hand, the cyclic predation within each partnership is responsible for the spiral waves observed inside each of these two types of spatial domains. We have shown, using square lattice simulations, that single-species spatial domains may be formed during the final stages of the collapse of three-species spatial domains and expand until they intersect a three-species spatial domain of the same partnership. We have investigated the conditions under which spherical spatial domains are able to grow, using both analytical arguments and numerical simulations, showing that there is a critical initial radius beyond which spherically symmetric spatial domains are expected to expand. We have investigated the corresponding impact on the average time evolution of the characteristic length scale of the spatial domains, identifying two different scaling regimes: a transient scaling regime, with a slower growth rate of the characteristic length scale $L$, takes place before the transition to a standard $L \propto t^{1/2}$ scaling law, resulting in an extended period of coexistence.

It is worth noticing that in previous work two different models, closely related to the one investigated in the present paper, have been investigated. The model considered in ref.  \cite{PhysRevE.96.012147} is similar to the one studied here, except for the inclusion of a unidirectional predation interaction, rather than a bidirectional one, between species $i$ and $i+1$. This leads to a completely different dynamical behaviour in which the evolution is never curvature dominated (the evolution does not follow the standard $L \propto t^{1/2}$ scaling law usually associated with a curvature dominated dynamics --- note that the authors of ref. \cite{PhysRevE.96.012147} have misidentified their model as being similar to the model V investigated in ref. \cite{Avelino-PRE-86-031119}). Unlike in our model, in the model presented in ref. \cite{PhysRevE.96.012147} an individual from the species $i$ is not able to select an individual from the species $i-1$. As a consequence, in that model large-scale coherent fluctuations of the interfaces may arise due to successive spiral wave fronts. This is in sharp contrast with the case studied in the present paper in which spiral waves cannot cross the interfaces between enemy partnerships spatial domains thus leading to much more localized fluctuations of the interfaces. On the other hand, in the model V studied in ref. \cite{Avelino-PRE-86-031119} the bidirectional predation interactions between species $i$ and $i+3$, which are present in the model studied in the present paper, are suppressed. In this case the impact on the dynamics is less significant (in comparison with that arising in the model considered in ref. \cite{PhysRevE.96.012147}). Still, the fact that the species of one partnership do not have predation interactions with all the species of the enemy partnership leads to the development of dynamical structures along the spatial domain interfaces and to the invasion of one-species spatial domains by neutral individuals of the enemy partnership. Hence, although the main effect which is responsible for the expansion of single-species spatial domains (the larger density of individuals in single-species compared to three-species spatial domains) is also present in the models studied in refs. \cite{Avelino-PRE-86-031119,PhysRevE.96.012147}, this effect is overshadowed by a more complex dynamics which prevents the growth of single-species spatial domains in those models. 

\begin{acknowledgments}
P.P.A. acknowledges the support by FEDER—Fundo Europeu de Desenvolvimento Regional funds through the COMPETE 2020—Operational Programme for Competitiveness and Internationalisation (POCI), and by Portuguese funds through FCT - Fundação para a Ciência e a Tecnologia in the framework of the project POCI-01-0145-FEDER-031938. J. M. acknowledges the support by NWO  - Netherlands Organisation for Scientific Research Visitor's Travel Grant 040.11.643. B.F.O. acknowledges Fundação Araucária, and INCT-FCx (CNPq/FAPESP) for financial and computational support. Funding of this work has also been provided by the FCT Grant No. UID/FIS/04434/2013.
\end{acknowledgments}

\bibliography{paper.bib}
\end{document}